\documentclass[aps,10pt,prl,twocolumn,showpacs]{revtex4-1}
\usepackage{amsmath,amssymb,amsfonts,amsthm}
\usepackage[ascii]{inputenc}
\usepackage{graphicx}
\usepackage[caption=false]{subfig}
\usepackage{bbm}
\usepackage[pdftex,bookmarks=false,colorlinks=true,linkcolor=blue,
citecolor=blue,filecolor=black,urlcolor=blue]{hyperref}

\graphicspath{{figures/}{../../calculations/analysis/plots/analysis_current_step_smalljump_t1024/}{../../theory/figures/}{../../../AHR/calculations/analysis/plots/analysis_AHR_current_step_t1024/}}

\begin{document}

\title{Searching for the Tracy-Widom distribution in nonequilibrium processes}


\author{Christian B. Mendl}
\email{mendl@stanford.edu}
\affiliation{
Stanford Institute for Materials and Energy Sciences, SLAC National Accelerator Laboratory, and
Geballe Laboratory for Advanced Materials,
Stanford University,
Stanford, CA 94305,
USA}

\author{Herbert Spohn}
\email{spohn@ma.tum.de}
\affiliation{
Zentrum Mathematik and Physik Department,\\
Technische Universit\"at M\"unchen,
Boltzmannstra{\ss}e 3,
85747 Garching,
Germany}

\date{May 13, 2016}

\begin{abstract}
While originally discovered in the context of the Gaussian Unitary Ensemble, the Tracy-Widom distribution also rules the height fluctuations of growth processes. This suggests that there might be other nonequilibrium processes in which the Tracy-Widom distribution plays an important role. In our contribution we study one-dimensional systems with domain wall initial conditions. For an appropriate choice of parameters the profile develops a rarefaction wave, while maintaining the initial equilibrium states far to the left and right, which thus serve as infinitely extended thermal reservoirs. For a Fermi-Pasta-Ulam type anharmonic chain we will demonstrate that the time-integrated current has a deterministic contribution, linear in time $t$, and fluctuations of size $t^{1/3}$ with a Tracy-Widom distributed random amplitude.
\end{abstract}

\maketitle

The famous Tracy-Widom (TW) distribution was discovered in the context of edge statistics of random matrices \cite{TracyWidom1993, TracyWidom1994}. To briefly recall, let us consider the specific case of a Gaussian Unitary Ensemble (GUE). One starts from $N \times N$ complex hermitian matrices, $A$, with probability density $Z^{-1} \exp[ - \frac{1}{2N} \mathrm{tr}A^2]$. Let $\lambda_1 < \cdots < \lambda_N$ denote the eigenvalues of $A$. Their spacings are of order $1$ and the largest eigenvalue has fluctuations governed by
\begin{equation}\label{1}
\lambda_N \simeq 2 N + N^{1/3}\xi_\mathrm{TW},
\end{equation}
asymptotically for large $N$. Here $\xi_\mathrm{TW}$ is a random amplitude with TW distribution. Its probability density function (pdf) has a typically asymmetric shape with mean $-1.771$, a left tail decaying as $\exp(-\tfrac{1}{12}\lvert s \rvert^3)$, and a right tail as $\exp(-\tfrac{4}{3}\lvert s \rvert^{3/2})$, see \cite{Bornemann2010} for definitions and numerical plots. As later established, the TW distribution also governs the edge statistics of Wigner matrices \cite{LY14} and sample covariance matrices \cite{LS14}. Thus it is regarded as the fixed point of a large universality class.

To a complete surprise, at the turn of the century it was discovered that the TW distribution also describes the shape fluctuations of a two-dimensional growing droplet in the KPZ universality class \cite{KPZ1986}. The first indication came from an exact solution of a particular growth model \cite{J00}. Subsequently it was understood that the TW distribution results whenever the macroscopic shape is curved, as confirmed by numerical simulations of a variety of growth models \cite{Alves2012,Ta12,Halpin-Healy2014} and experimental studies \cite{MM05, TakeuchiSanoPRL2010, TaSa12, YunkerPRL2013}, see \cite{K97, BaSt95, FeSp11, HaTa15} for reviews.

With this background information one may wonder whether there are further dynamical non-equilibrium processes with fluctuations governed by TW. Our search starts from the observation that if the height evolves according to the one-dimensional KPZ equation, then its slope, $u$, is governed by the stochastic Burgers equation
\begin{equation}\label{4a}
\partial_t u + \partial_x \big(u^2 -D\partial_x u + \xi(x,t)\big) = 0,
\end{equation}
which we read as a stochastic conservation law. Its current has three generic features consisting of (i) a nonlinear systematic current, as $u^2$, (ii) a dissipative term, here the gradient term $-D\partial_x u$ with diffusion constant $D$, and (iii) a space-time random current with short range correlations, $\xi(x,t)$, which models all fast degrees of freedom. Curved initial data for the KPZ equation correspond to domain wall initial conditions for the Burgers equation with the property that the solution develops a rarefaction wave. The height corresponds to the time-integrated current. This perspective suggests to look for other one-dimensional systems with conservation laws. Leaving quantum models aside, an obvious and physically interesting candidate is a one-dimensional classical fluid of interacting point particles governed by Newton's equations of motion. In analogy the initial state should be domain wall, i.e., right and left half line are in a thermal state with distinct equilibrium parameters, chosen such that a rarefaction wave is generated.

A one-dimensional fluid has three conservation laws. Thus the real challenge is to go from the well understood scalar equation \eqref{4a} to its vector valued version, known as nonlinear fluctuating hydrodynamics in one dimension \cite{vanBeijeren2012, Spohn2014}. Previous studies considered small deviations from a homogeneous equilibrium state \cite{Dhar2015,Straka2014,MS14,MS15,Spohn2015}. In our contribution we investigate the time evolution of such strongly interacting classical many-body systems starting from a non-equilibrium domain wall state with the goal to uncover the TW distribution.

From earlier experience \cite{MS15}, the TW distribution can be meaningfully numerically validated by having on the order of $10^7$ samples available. For a fluid with short range interactions such a task is challenging and, as generally adopted, we simplify to Fermi-Pasta-Ulam type anharmonic chains. The chain dynamics has to be sufficiently chaotic and molecular dynamics is currently the only available tool to observe the TW distribution. To properly set up the numerical experiment we need however the guidance from the mesoscopic theory of nonlinear fluctuating hydrodynamics \cite{Spohn2014}. This theory applies also to stochastic systems. Thus, before tackling the real problem, it is instructive to first study stochastic lattice gases, which will serve as a blueprint. We recall the well understood case of a single conserved field. Turning to stochastic models with two conserved fields, one already enters unexplored territory.

Let us then first consider a one-dimensional lattice gas with Kawasaki dynamics. There is at most one particle per site. The occupation variables will be denoted by $\eta_j $, $j \in \mathbb{Z}$, and take only the values $0,1$. The dynamics proceeds via independent nearest neighbor exchanges. For the bond $(j,j+1)$ the exchange rate is denoted by $c_{j,j+1}(\eta)$, which is assumed to be strictly positive and of finite range. Then the particle number is the only conserved field and for every choice of the average density $\rho$, $0 \le \rho \le 1$, there is a unique steady state $\langle \cdot\rangle_{\rho}$. The steady state current is given through $\mathsf{j}(\rho) = \langle c_{j,j+1}(\eta)(\eta_j - \eta_{j+1}) \rangle_\rho$. On a macroscopic scale, ignoring tiny fluctuations, the mass density $\rho(x,t)$ is then governed by the deterministic conservation law
\begin{equation}\label{5}
\partial_t \rho(x,t) + \partial_x \mathsf{j}\big(\rho(x,t)\big) = 0.
\end{equation}

The domain wall initial state consist of $\langle \cdot\rangle_{\rho_-}$ in the left half lattice and $\langle \cdot\rangle_{\rho_+}$ in the right half lattice with parameters $\rho_- \neq \rho_+$. Our interest are the statistical properties of the current integrated along the ray $\{x = \mathsf{v} t, t \geq 0\}$. We first have to figure out the leading deterministic contribution, which is obtained by solving Eq. \eqref{5} with initial condition $\rho(x,0) = \rho_{-}$ for $x \le 0$ and $\rho(x,0) = \rho_{+}$ for $x > 0$. Its solution is self-similar of the form $\rho(x,t) = \rho_{\mathrm{dw}}(x/t)$. Along the ray $x = \mathsf{v} t$, the density is constant and given by $\rho_{\mathsf{v}} = \rho_{\mathrm{dw}}(\mathsf{v})$. The construction of $\rho_{\mathrm{dw}}$ is discussed, for example, in \cite{HoldenRisebro}. If $\rho_{-} < \rho_{+}$, then $\rho_{\mathrm{dw}}(x) = \rho_{-}$ for $x \le x_{-}$ and $\rho_{\mathrm{dw}}(x) = \rho_{+}$ for $x_{+} \le x$. For $x_{-} \le x \le x_{+}$, $\rho_{\mathrm{dw}}$ has up-ward jumps, the shock discontinuities, and smooth, strictly increasing segments, known as rarefaction waves. The for us crucial point is that the TW distribution can be detected only if the observation ray lies inside a rarefaction wave. In particular we require $x_- < \mathsf v < x_+$ and $\mathsf{v}t$ not to coincide with a shock location. It may happen that $\rho_{\mathrm{dw}}$ consists of a single shock. For such initial domain wall states, there will be no TW distribution.

We still have to define the time-integrated current, which becomes more transparent in the $x,t$ continuum setting. But our conclusion holds with the obvious correspondence for a spatial lattice. Let us assume that one particular realization of the conserved field, $\mathfrak{u}(x,t)$, and its current, $\mathfrak{j}(x,t)$, are linked through the conservation law
\begin{equation}\label{6}
\partial_t \mathfrak{u}(x,t) + \partial_x \mathfrak{j}(x,t) = 0.
\end{equation} 
Thus, as a property special for one dimension, the vector field $(-\mathfrak{u}, \mathfrak{j})$ is curl-free and admits a potential, $\Phi(x,t)$, up to a constant which we fix by $\Phi(0,0) = 0$. A particular definition, to be used below, is
\begin{equation}\label{7}
\Phi(x,t) = \int _0^t dt'\, \mathfrak{j}(x,t') - \int_0^x dx' \mathfrak{u}(x',0).
\end{equation} 
In the KPZ context, $\Phi(x,t)$ would be the height function. Applying the KPZ scaling theory \cite{KMH92,S12} to our stochastic lattice gas, if the observation ray lies inside a rarefaction wave and if $\mathsf{j}''(\rho_{\mathsf{v}}) \neq 0$, then
\begin{equation}\label{8}
\Phi(\mathsf{v} t, t) \simeq a_0 t + (\Gamma t)^{1/3} \xi_\mathrm{TW}
\end{equation}
for large $t$, in complete analogy to \eqref{1}. Here $a_0$ and $\Gamma$ are model-dependent parameters. The asymptotics \eqref{8} has been proved for the totally asymmetric simple exclusion process \cite{J00} and its partially asymmetric version \cite{TracyWidom2009}.

\begin{figure}[!t]
\centering
\subfloat[]{\includegraphics[width=0.5\columnwidth]{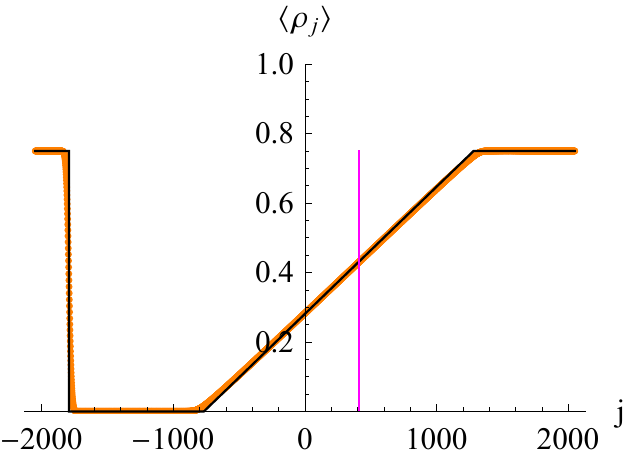}}
\subfloat[]{\includegraphics[width=0.5\columnwidth]{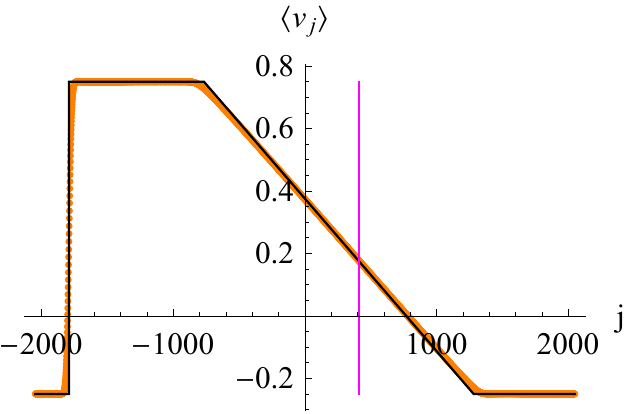}}
\caption{Density (a) and velocity (b) profile at time $t = 1024$ of the Leroux model with domain wall initial conditions. The orange dots are molecular dynamics results and the black line shows the theoretically predicted profile. The jump on the left is the shock.}
\label{fig:AHR_profile}
\end{figure}

Next we extend our discussion to stochastic lattice gases with \emph{two components}. Rather then attempting to deal with the general case we study the Leroux stochastic lattice gas, which is a particular case of the AHR model \cite{AHR98,FerrariSasamotoSpohn2013}. The occupation variables $\eta_j$ take the values $-1,0,1$. The ${-1}$-particles jump only to the left, while the $1$-particles jump only to the right, both with rate $1$. At a collision, an adjacent pair of $1|{-1}$ exchanges to ${-1}|1$ with rate $2$. The number of $1$'s and ${-1}$'s are conserved and our particular choice of rates implies that in the steady state the $\eta_j$'s are independent. For the macroscopic description it is more convenient to switch to the number of holes $\rho = 1 - \rho_{1} - \rho_{-1}$ and the velocity $v =\rho_{1} -\rho_{-1}$, with $\rho_{\sigma}$ the density of $\sigma$-particles. Then the two macroscopic currents are $j_{\rho} = -\rho v$ and $j_{v} = 1 - (\rho + v^2)$. A domain wall initial state is defined by $\rho(x) = \rho_{-}$, $v(x) = v_{-}$ for $x < 0$ and $\rho(x) = \rho_{+}$, $v(x) = v_{+}$ for $x > 0$. The solution of the Euler equations with this initial condition is again self-similar, but its construction is more complicated than for the scalar case. In the mathematical literature this is known as Riemann problem \cite{Wendroff1972,MenikoffPlohr1989,Bressan2013}. The construction for the Leroux system is discussed in \cite{MS16}. To our surprise, the solutions to the Rankine-Hugoniot conditions coincide with the rarefaction integral curves except for their orientation \cite{MS16}, which is the signature of the Temple class \cite{T82}. The rarefaction profiles are predicted to be linear. In our Monte-Carlo simulation we consider a ring of $L = 4096$ lattice sites. In one half lattice we distribute particles in their steady state with parameters $\rho_-,v_-$ and in the other half lattice with parameters $\rho_+,v_+$. Note that now there are two domain walls, one at $0$ and the other one at $L/2$. As shown in Fig.~\ref{fig:AHR_profile}, the Monte-Carlo simulation confirms very precisely the predicted profiles. Note that the shock has a width of a few lattice spacings only. The quantity of interest is the time-integrated current along a ray inside the rarefaction wave, as indicated by the purple line in Fig.~\ref{fig:AHR_profile}.

\begin{figure}[!t]
\centering
\subfloat[]{\includegraphics[width=0.5\columnwidth]{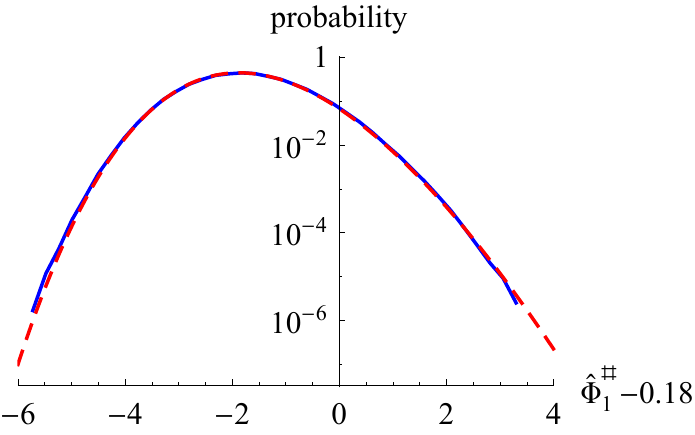}\label{fig:AHR_distribution_TW}}
\subfloat[]{\includegraphics[width=0.5\columnwidth]{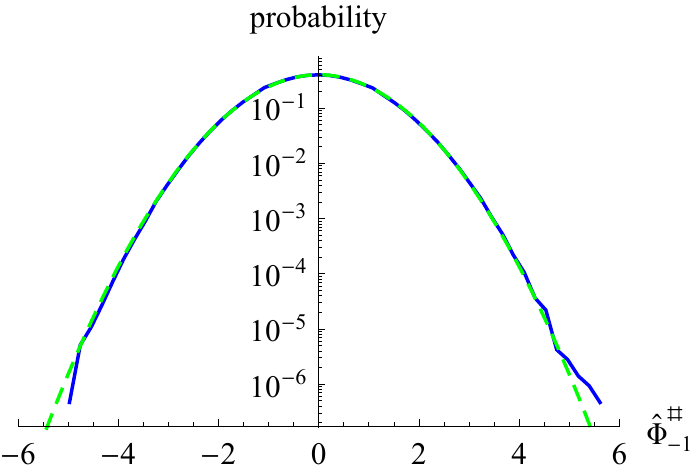}\label{fig:AHR_distribution_Gauss}}
\caption{(a) Pdf of $(\Gamma_1 t)^{-1/3}\Phi^{\sharp}_1(t)$ for the Leroux model at $t = 1024$ compared with the Tracy-Widom distribution (red dashed) and (b) pdf of $(1.34 t)^{-1/2}\,\Phi^{\sharp}_{-1}(t)$ compared with a normalized Gaussian (green dashed).}
\end{figure}

There is however one difficulty which we did not anticipate so far. The microscopic current has two components. In Fig.~\ref{fig:AHR_distribution_TW} we show the pdf for one very specific linear combination, $\Phi^{\sharp}_1(t)$, of these currents. Nonlinear fluctuating hydrodynamics predicts the scale $(\Gamma_1 t)^{1/3}$ with $\Gamma_1 = 0.539$. On that scale TW fluctuations are confirmed with high accuracy, except for a shift of $0.18$. For the shorter time $t = 512$, the fit is equally accurate but the shift equals $0.27$, and thus we expect the shift to vanish for longer times. On the other hand measuring a distinct linear combination, denoted by $\Phi^{\sharp}_{-1}(t)$, results in central limit type fluctuations of size $(1.34 t)^{1/2}$, as is shown in Fig.~\ref{fig:AHR_distribution_Gauss}. To understand the mechanism at work, one has to enter more deeply into nonlinear fluctuating hydrodynamics.

We now have the tools to analyse a one-dimensional fluid with domain wall initial conditions. The particles have mass $1$, positions $q_j$, and momenta $p_j$. As a commonly adopted simplification, we restrict to nearest neighbor interactions, which amounts to an anharmonic chain with the equations of motion
\begin{equation}\label{9}
\ddot{q}_j = V'(q_{j+1}- q_j) - V'(q_j - q_{j-1}).
\end{equation}
It is convenient to regard \eqref{9} as a discrete nonlinear wave equation with displacements $q_j \in \mathbb{R}$, see \cite{Spohn2014} for more details. The potential $V$ can be rather general, but at least $V(x) \to \infty$ for either $x \to \infty$ or $x \to -\infty$. Disregarding integrable exceptions, like harmonic and Toda chains, the wave equation has three locally conserved fields, namely stretch $r_j = q_{j+1} - q_j$, momentum $p_j = \dot{q}_j$, and energy $e_j = \tfrac{1}{2}p_j^2 + V(r_j)$. As follows directly from the equations of motion, the respective microscopic currents are ${\big({-p_j}, -V'(r_{j-1}), -p_j V'(r_{j-1})\big)}$. Thermal equilibrium is described by the canonical ensemble with pressure $P$, average momentum $\nu$, and inverse temperature $\beta$. We introduce the average stretch by $r(P,\beta) = {Z^{-1} \int dx\,x \exp[- \beta(V(x) + P x)]}$ and the average internal energy by $e(P,\beta) = (2\beta)^{-1} + {Z^{-1}\int dx\,V(x)\exp[- \beta(V(x) + Px)]}$, with inverses ${P = P(r,e)}$, ${\beta = \beta(r,e)}$. 

Initially we impose a domain wall state, for which the left half-lattice is in a state of thermal equilibrium with parameters $\vec{u}_{-}$ and the right half-lattice with parameters $\vec{u}_{+}$, where we combined the average value of the physical fields as $3$-vector ${\vec{u} = (u_1,u_2,u_3) = (r,\nu,\mathfrak{e})}$, $\mathfrak{e} = e + \tfrac{1}{2}\nu^2$ the total energy. Following \eqref{7}, for such random initial data we determine component-wise the microscopic time-integrated current $\Phi_\alpha (\mathsf{v}t,t)$, $\alpha = 1,2,3$. Based on the insight from the Leroux model, for an appropriate choice of $\mathsf{v}$ and a carefully adjusted linear combination of the $\Phi_\alpha(\mathsf{v} t, t)$'s, one should find a long-time behavior as in \eqref{8}.

As before the macroscopic Euler currents are obtained as (local) thermal average of the microscopic currents, i.e., ${\big({-\langle p_j \rangle_{P,\nu,\beta}}, -\langle V'(r_{j-1}) \rangle_{P,\nu,\beta} -\langle p_j V'(r_{j-1})\rangle_{P,\nu,\beta} \big)}$. Hence 
\begin{equation}\label{10}
\vec{\mathsf{j}}(r,\nu, \mathfrak{e}) = \big({-\nu}, P(r, \mathfrak{e} - \tfrac{1}{2}\nu^2), \nu P(r, \mathfrak{e} - \tfrac{1}{2}\nu^2)\big).
\end{equation}
The macroscopic conservation laws then read
\begin{equation}\label{12}
\partial_t u_{\alpha}(x,t) + \partial_x \mathsf{j}_\alpha(\vec{u}(x,t)) = 0,
\end{equation}
$\alpha = 1,2,3$. One has to solve the Riemann problem with initial conditions
\begin{equation}\label{13}
\vec{u}(x,0) = \vec{u}_{-}\ \ \text{for} \ \ x \le 0, \quad \vec{u}(x,0) = \vec{u}_{+}\ \ \text{for} \ \ x > 0.
\end{equation} 
 Its solution is self-similar as $\vec{u}(x,t) = \vec{u}_{\mathrm{dw}}(x/t)$, but the construction of $\vec{u}_{\mathrm{dw}}$ is much more involved than in the case of a single conservation law. We refer to the most instructive and readable exposition \cite{Bressan2013}. Our Riemann problem is rather close to the one for one-dimensional fluids, see the discussion in \cite{MenikoffPlohr1989} and the pioneering work of Bethe \cite{Bethe1942}. The analysis is based on the linearization of \eqref{12}, which is determined by the $3 \times 3$ matrix
\begin{equation}\label{14}
A(\vec{u}) =
\begin{pmatrix}
0 & -1 & 0 \\
\partial_r P & -\nu \partial_e P & \partial_e P \\
\nu \partial_r P & \,P - \nu^2\,\partial_e P\, & \nu \partial_e P \\
\end{pmatrix}.
\end{equation}
$A(\vec{u})$ has the eigenvalues $0, \pm c(\vec{u})$ with $c(\vec{u})$ the adiabatic speed of sound, $c^2 = -\partial_r P + P\,\partial_e P > 0$. The eigenvalue $0$ yields a contact discontinuity, while the eigenvalues $\pm c$ govern the rarefaction waves.

Even if the solution to the Riemann problem could be found, we face a further difficulty. Numerically the lattice is finite with periodic boundary conditions. So, in fact, one has to solve the \emph{periodic} Riemann problem in the periodic box $[-\tfrac{L}{2}, \tfrac{L}{2}]$. Concretely, one solves the Riemann problem with $\vec{u}_{-}\vert\vec{u}_+$ centered at $0$ and in addition the one with $\vec{u}_{+}\vert\vec{u}_{-}$ centered at some $\ell_0$, $\lvert\ell_0\rvert \le \tfrac{L}{2}$. Thereby the desired self-similar ring solution is obtained, however limited in time by the first collision between shocks or rarefaction waves. 

So far we did not specify the potential $V$. Based on previous experience \cite{MS14,MS15}, we choose the hard-point potential $V(x) = 0$ for $x > 0$ and $V(x) = \infty$ for $x\leq 0$. Another popular choice would be the FPU potential $V(x) = \tfrac{1}{2}x^2 + \tfrac{1}{3}ax^3 + \tfrac{1}{4}bx^4$, which however is known to have slower relaxation \cite{Dhar2015}. To ensure deterministic chaos the standard practice is to adopt alternating masses $m_{2j} = m_0$ and $m_{2j+1} = m_1$ with mass ratio $\kappa = m_1/m_0$. In our simulations we set $m_0 = 1$ and $\kappa = 3$, which seems to provide sufficiently strong dynamical mixing. Since the potential is hard core of zero diameter, the thermodynamics is the one of an ideal gas, i.e., $P = 2e/r$, $\beta = 1/(2e)$, and $c = r^{-1}\sqrt{12e/(1 +\kappa)}$. To our advantage the Riemann problem can still be solved in closed form \cite{MS16}.

\begin{figure}[!ht]
\centering
\includegraphics[width=0.75\columnwidth]{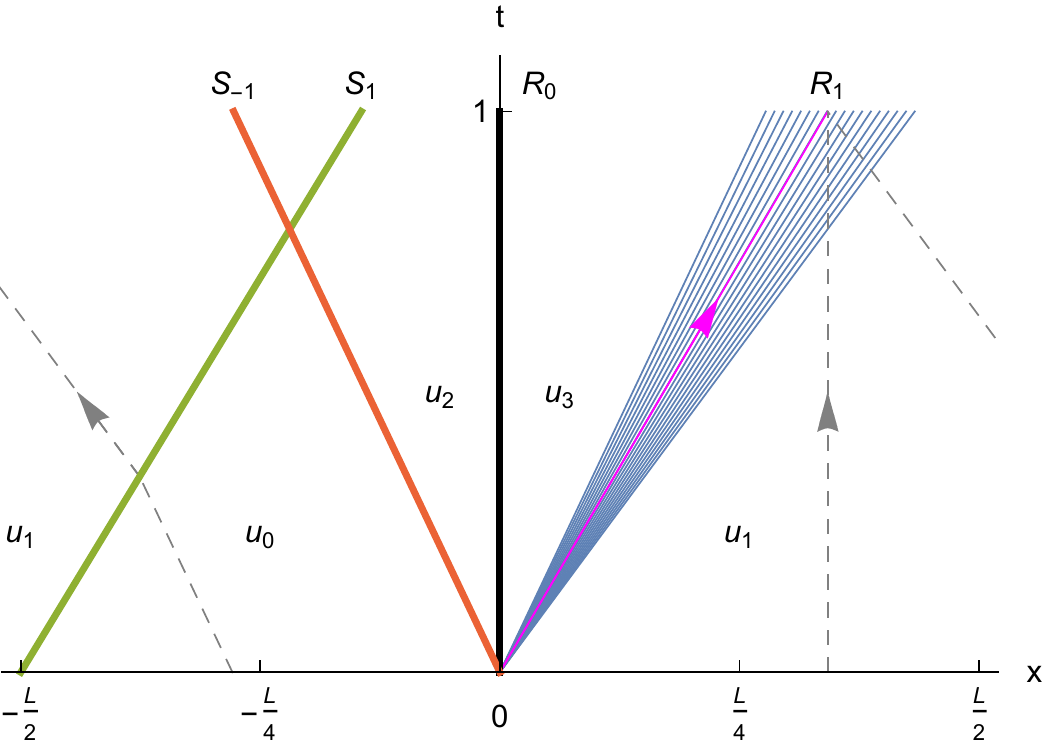}
\caption{Analytical solution of the periodic Riemann problem for an anharmonic chain of hard-point particles with alternating masses.}
\label{fig:Riemann_analytic}
\end{figure}

One still has a considerable freedom in the choice of the initial $\vec{u}_{\mathrm{\pm}}$ and $\ell_0$. To arrive at an optimal rarefaction wave, we varied its size, its gradient, and the time of first collision. Our preferred choice is shown in Fig.~\ref{fig:Riemann_analytic}, for a more complete discussion we refer to \cite{MS16}. In the white areas the self-similar solution $\vec{u}_{\mathrm{dw}}(x/t)$ has a constant value. We parametrize as $\vec{u} = (r, \nu, e)$. Then $\vec{u}_1 = (1,0,1)$, $\beta_1 = \tfrac{1}{2}$, $P_1 = 2$ and $\vec{u}_3 = (\frac{5}{4}, -\frac{2}{5} \sqrt{3}, \frac{16}{25})$, $\beta_3 = \frac{25}{32}$, $P_3 = \frac{128}{125}$. The rarefaction wave, $R_1$, smoothly interpolates between these values. $R_1$ is constructed from the eigenvector of $A(\vec{u})$ with eigenvalue $c$, hence the index $1$. $S_1$ is a shock curve with discontinuity from $\vec{u}_1$ to $\vec{u}_0$. The contact discontinuity, $R_0$, is associated with the eigenvalue $0$. At $R_0$ the profile jumps from $\vec{u}_2$ to $\vec{u}_3$. Finally, $S_{-1}$ is another shock curve, which crosses $S_1$ before the largest recorded time, but lies well away from the purple observation ray in the interior of $R_1$. The jumps at $R_0$, $S_1$ and $S_{-1}$ are the price for imposing periodic boundary conditions. The for us relevant part is the rarefaction wave $R_1$, which is maintained by the two bordering thermal states with parameters $\vec{u}_3$ and $\vec{u}_1$. As a general fact for anharmonic chains, the entropy remains constant across the rarefaction fan.

\begin{figure}[!ht]
\centering
\subfloat[]{\includegraphics[width=0.5\columnwidth]{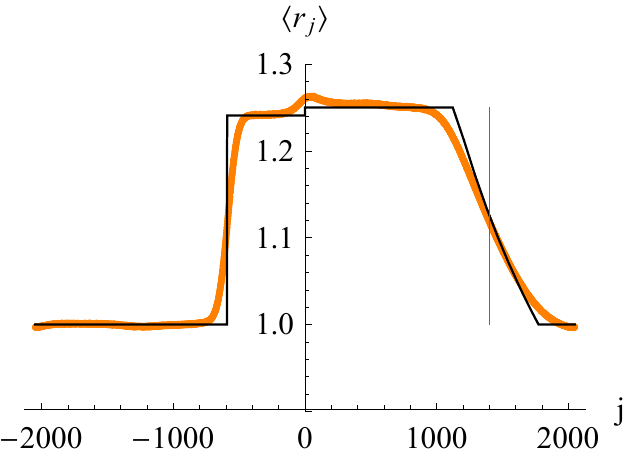}}
\subfloat[]{\includegraphics[width=0.5\columnwidth]{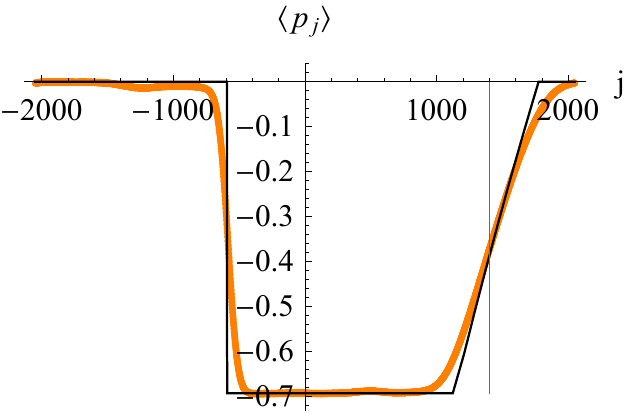}}
\caption{Average profile of the stretch $r_j(t)$ and momentum $p_j(t)$ at $t = 1024$. Orange dots show numerical molecular dynamics results, and the black curve the analytical solution of the Riemann problem. The purple vertical line marks $\mathsf{v} t$.}
\label{fig:r_profile}
\end{figure}

For the molecular dynamics simulation we choose a system size of $L = 4096$ and time $t= 1024$. In $[-\tfrac{L}{2}, 0]$ we prepare thermal equilibrium with parameters $\vec{u}_0$ and in $[0,\tfrac{L}{2}]$ with parameters $\vec{u}_1$, which just means to sample the initial $r_j,p_j$ independently according to these parameters. In Fig.~\ref{fig:r_profile} we show the numerical profiles for $\langle r_{j}(t)\rangle$ and $\langle p_{j}(t)\rangle$. The rarefaction wave is smooth. The current is integrated along the ray $\{x = \mathsf{v} t, t \ge 0\}$ with the particular choice $\mathsf{v} = \frac{64}{27\sqrt{3}}$, as indicated by the purple line in Figs.~\ref{fig:Riemann_analytic}, \ref{fig:r_profile}. Along this observation ray the equilibrium parameters are constant and denoted by $\vec{u}_{\mathsf{v}}$. In our case $\vec{u}_{\mathsf{v}} = (\frac{9}{8}, -\frac{2}{3 \sqrt{3}}, \frac{64}{81})$ satisfying $c(\vec{u}_{\mathsf{v}}) = \mathsf{v}$. 

To predict the current fluctuations, one has to follow the propagation of small perturbations relative to the deterministic background profile. Let us denote by $\langle\cdot|\cdot\rangle$ the scalar product for $3$-vectors and by $\tilde{\psi}_\sigma$ the left eigenvectors of $A(\vec{u}_{\mathsf{v}})$, $\tilde{\psi}_\sigma A(\vec{u}_{\mathsf{v}}) = \sigma \mathsf{v} \tilde{\psi}_\sigma$, $\sigma = 0,\pm 1$. The projected current components, with their asymptotic value subtracted, are then defined by
\begin{equation}
\Phi^{\sharp}_{\sigma}(t) = \big\langle \tilde{\psi}_{\sigma} \vert\vec{\Phi}(\mathsf{v}t,t) - t (\vec{\mathsf{j}}(\vec{u}_{\mathsf{v}}) - \mathsf{v} \vec{u}_{\mathsf{v}})\big\rangle.
\end{equation}
Now, $\Phi^{\sharp}_{1}(t)$ results from perturbations propagating with velocity $\mathsf{v}$ along the observation ray. As for a single component, they will build up anomalous fluctuations and for the long time asymptotics one expects that
\begin{equation}\label{18}
\Phi^{\sharp}_1(t) \simeq (\Gamma_1 t)^{1/3} \xi_\mathrm{TW}.
\end{equation}
On the other hand the components $\sigma = -1,0$ are linked to perturbations which propagate as indicated by the gray dashed arrowed lines in Fig.~\ref{fig:Riemann_analytic}. Their time integral is sampled from essentially independent space-time regions implying a central limit type behavior as
\begin{equation}\label{19}
\Phi^{\sharp}_{\sigma}(t) \simeq (\Gamma_\sigma t)^{1/2}\xi_\mathrm{G},\qquad \sigma = 0, -1,
\end{equation}
where $\xi_\mathrm{G}$ is a normalized Gaussian random variable. Therefore the joint pdf of $\vec{\Phi}(\mathsf{v}t,t)$ is predicted to have a width of order $t^{1/3}$ along the direction defined by $\tilde{\psi}_1$, while along all other directions the width grows as $t^{1/2}$ asymptotically. The value of the non-universal coefficient $\Gamma_1$ can be derived in the framework of nonlinear fluctuating hydrodynamics. On the other hand there is no estimate available for $\Gamma_0$ and $\Gamma_{-1}$.

To have the observation ray inside a rarefaction wave is a robust feature. However to single out the TW distribution one still has to pick the appropriate linear combination of the time-integrated currents, as noted already for the Leroux stochastic lattice gas.
\begin{figure}[!ht]
\centering
\subfloat[]{\includegraphics[width=0.5\columnwidth]{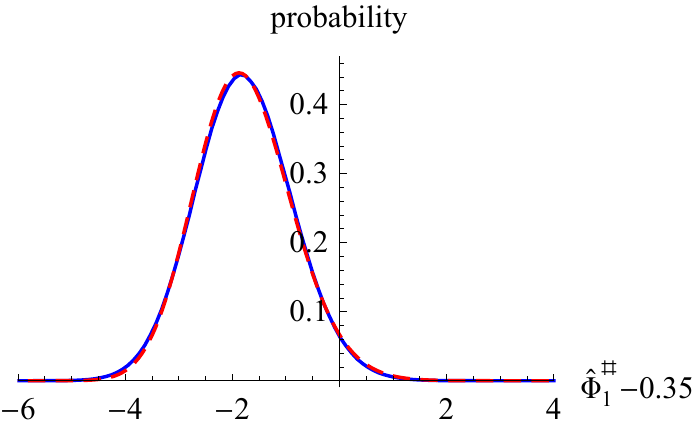}}
\subfloat[]{\includegraphics[width=0.5\columnwidth]{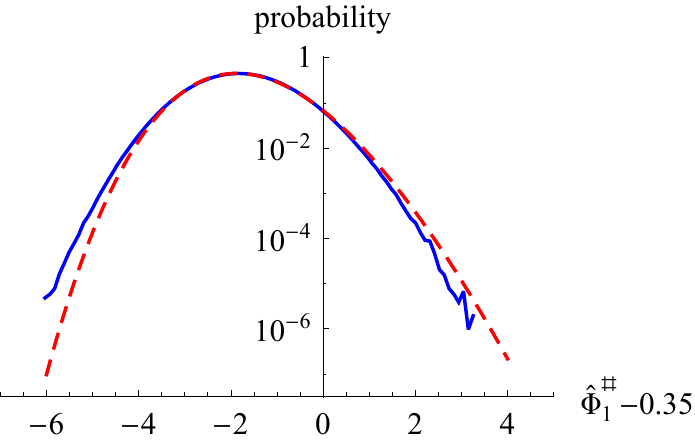}\label{fig:hardpoint_phi_distribution_TW_log}}\\
\subfloat[]{\includegraphics[width=0.5\columnwidth]{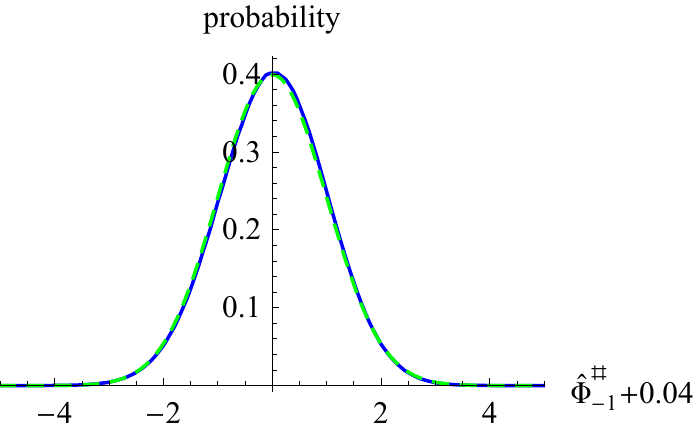}}
\subfloat[]{\includegraphics[width=0.5\columnwidth]{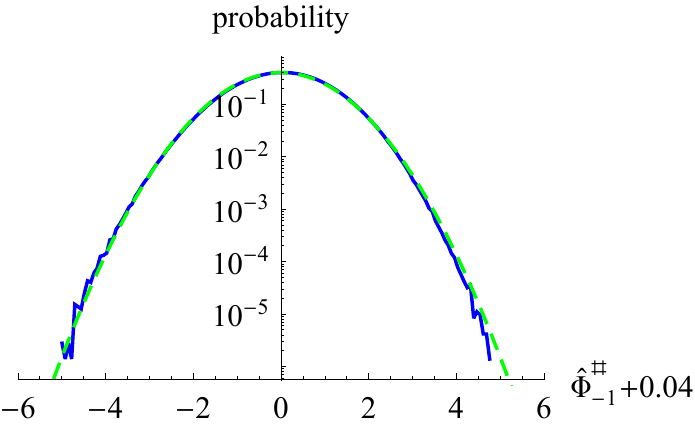}}
\caption{(a), (b) Statistical distribution of $\hat{\Phi}^{\sharp}_1(t) = (\Gamma_1 t)^{-1/3}\,\Phi^{\sharp}_1(t)$ obtained from molecular dynamics simulations at $t = 1024$. The red dashed curve is the Tracy-Widom pdf. (c), (d) The projection $(\Gamma_{-1} t)^{-1/2}\,\Phi^{\sharp}_{-1}(t)$ closely follows a Gaussian distribution (green dashed).}
\label{fig:hardpoint_phi_distribution}
\end{figure}
In Fig.~\ref{fig:hardpoint_phi_distribution} we show the pdf for the time-integrated current for the projections \eqref{18} and \eqref{19}. The qualitative agreement is surprisingly accurate, only the tails in Fig.~\ref{fig:hardpoint_phi_distribution_TW_log} slightly deviate from the predicted TW distribution. The mean is fitted by the value indicated at the label of the $1$-axis, which is expected to vanish in the long time limit. We measure $\Gamma_1 = 0.86$ as a numerical fit, while the theory value is $0.559$. As further test of the theoretical prediction \eqref{18} and \eqref{19}, we have measured the standard deviation of $\Phi^{\sharp}_{\sigma}(t)$ for various time points, and indeed observe $\sim t^{1/3}$ scaling for $\sigma = 1$ and $\sim t^{1/2}$ for $\sigma = 0, -1$ (data not shown).

In summary, through molecular dynamics simulations we have established that the statistics of time-integrated currents of a particular classical anharmonic chain is governed by the TW distribution on the time scale $t^{1/3}$. Implicitly we have used that the chain dynamics is chaotic, which is expected to hold generically away from the regime of very low temperatures. The TW distribution is observed under two conditions: the current has to be integrated along a ray located inside a rarefaction wave and to be projected along the respective eigenvector of the linearized Euler equation. We expect that such a result is robust and generically holds also for other anharmonic chains and for one-dimensional fluids. Based on nonlinear fluctuating hydrodynamics, the crucial condition is to have Euler currents which depend nonlinearly on the conserved fields. Under this condition, the TW distribution is expected to turn up for time-integrated currents also in quantum spin chains and one-dimensional quantum fluids with domain wall initial conditions.
\begin{acknowledgments}
\textbf{Acknowledgements} The work of HS has been supported through a senior chair of the Fondation Sciences Math\'{e}matiques de Paris. HS greatly benefited from discussions with O.~Glass and S.~Olla on hyperbolic conservation laws. CM acknowledges support from the Alexander von Humboldt Foundation and computing resources of the Leibniz-Rechenzentrum.
\end{acknowledgments}


\end{document}